\documentclass[journal,article,submit,pdftex,moreauthors]{Definitions/mdpi} 

\firstpage{1} 
\pubvolume{1}
\issuenum{1}
\articlenumber{0}
\pubyear{2024}
\copyrightyear{2024}
\datereceived{ } 
\daterevised{ } 
\dateaccepted{ } 
\datepublished{ } 
\hreflink{https://doi.org/} 
\renewcommand\linenumbers{}

\Title{Sex Differences in Hierarchical and Modular Organization of Functional Brain Networks: Insights from Hierarchical Entropy and Modularity Analysis}

\TitleCitation{Title}

\Author{Wenyu Chen $^{1,}$\orcidA{},Ling Zhan $^{1,}$\orcidC{}, and Tao Jia $^{1,}$*\orcidB{}}

\AuthorNames{Firstname Lastname, Firstname Lastname and Firstname Lastname}

\address{%
$^{1}$ \quad College of Computer and Information Science, Southwest University, Chongqing, 400715, P. R. China\\
}

\corres{Correspondence: tjia@swu.edu.cn}

\abstract{Existing studies have demonstrated significant sex differences in the neural mechanisms of daily life and neuropsychiatric disorders.
The hierarchical organization of the functional brain network is a critical feature for assessing these neural mechanisms.
But sex differences on the hierarchical organization is not fully investigated. 
Here, we explore whether hierarchical structure of brain network differs between females and males.
At the group level, we measure the hierarchical entropy and the maximum modularity of each individual, and identify a significant negative correlation between the complexity of hierarchy and modularity in brain networks.
On average, female brain networks have stronger connectivity within the module, whereas male brain networks demonstrate more complex hierarchy.
At the consensus level, we use co-classification matrix to investigate the detailed differences in the hierarchical organization between sexes and observe that 
females and males exhibit different interaction patterns of brain regions in the dorsal attention network (DAN) and visual network (VIN). 
Our finding suggests that females and males employ different network topology to achieve brain functions. 
In addition, the negative correlation between hierarchy and modularity implies a need to balance the complexity in hierarchical organization of the brain network, which shed light on future studies of brain functions.
}


\begin{document}

\section{Introduction}
Biological sex is a crucial variable in the field of neuroscience because it relates both the neural underlying daily life and the incidence of neuropsychiatric disorders \citep{beery2011sex, pawluski2020sex, hawkes2024sex}.
For instance, females generally exhibit more activation in the prefrontal regions implicated in affective processing that may result in better performance on emotion regulation \citep{nolen2012emotion, mak2009sex} and higher susceptibility to emotional disorders such as anxiety and depression \citep{campbell2006acceptability,garnefski2004cognitive,joormann2010emotion,deckert2020subjective,luo2022accelerated}.
In contrast, males typically display higher synchronizations in brain regions such as bilateral posterior cingulate and left middle
occipital gyrus, which contribute to the better performance on mental rotation task  \citep{cimadevilla2020spatial}, but may give rise to higher risk for autism spectrum disorder (ASD) \citep{lawrence2020sex} and attention deficit hyperactivity disorder (ADHD) \citep{rosch2018adhd}.
A neural basis understanding of sex differences in healthy adults may provide further explanations on daily life activities and neuropsychiatric disorders.

Recently, advanced neuroimaging techniques like functional magnetic resonance imaging (fMRI) \citep{logothetis2001neurophysiological} and theories like network science have emerged as powerful tools for elucidating the sex differences in the topological properties of functional brain network \citep{bullmore2009complex,bassett2017network}.
Many studies on sex differences start to look into the functional connectivity of the brain network and find substantial differences in connection strength  \citep{satterthwaite2015linked,zhang2018functional,ritchie2018sex}. 
While using the connection strength is straightforward, it only reflect a very simple perspective of the network. 
To further explore the functional brain network, topological properties, such as hierarchical structure and modular structure of network, becomes points of interests \citep{bullmore2012economy}. Several studies report stronger intra-module connections in females and stronger inter-module connections in males \citep{satterthwaite2015linked,ritchie2018sex}, implying that males and females may have different hierarchical organizations in the brain network.

Modules are defined as highly connected clusters of entities that are sparsely connected to entities outside. They are believed to come from the hierarchy of the network because a pure random network does not exhibit modular structures \citep{barabasi2016network}. The hierarchy and modularity offer new perspectives to exploring brain functions across various scales \citep{bassett2010efficient, sporns2016modular, betzel2017multi}.
For example, research utilizing features reveals specific age-related effects on the segregation of modules within brain networks \citep{betzel2014changes,betzel2015functional}.
Several studies have highlighted that this type of hierarchy significantly enhances the brain’s ability to process complex information, and greatly improves efficiency, robustness, and adaptability \citep{bassett2010efficient, meunier2010modular, meunier2009hierarchical}. 
However, current research primarily focuses on exploring modularity or hierarchy \citep{sporns2016modular,esfahlani2021modularity} with little attention given to their inter-relationships. Investigating these relationships could enhance our comprehensive understanding of brain network topology across various scales.  Moreover, studies on sex differences predominantly concentrate on modularity \citep{ingalhalikar2014sex, satterthwaite2015linked,fauchon2021sex}, lacking detailed analysis of the differences in hierarchy between female and male brain networks.

In addition, current studies on sex differences usually focus on the positively correlated brain network, with only a few exceptions considering the negatively correlated brain network \citep{stumme2020functional}.
But recent findings have highlighted the pathophysiological significance of negative connections in the resting state, particularly between the default mode network and the dorsal attention network \citep{li2021co,mitra2018principles,baldassarre2014large,ramkiran2019resting}. Positive connections during rest may signify functional integration, whereas negative connections could denote functional segregation \citep{fox2005human, spreng2016attenuated}. Positive and negative connections probably play equally important role in brain functions. Therefore, the functional brain network with both positive and negative strength should be considered for a more thorough investigation of the sex difference.

In this study, we investigate the correlations between hierarchy and modularity in brain networks that includes positive and negative connections, and elucidate sex-based differences in hierarchical and modular structures. 
Specifically, we employ hierarchical clustering to generate dendrograms for each individual. We calculate the hierarchical entropy that reflects the complexity of hierarchical structures, and the maximum modularity that is associated with the connection strength within the optimal partition. We identify a negative correlation between the hierarchical entropy and the maximum modularity. We find significant difference between males and females. On average, male brain networks demonstrate more complex hierarchy, whereas female brain networks have stronger connectivity within the module. 
Moreover, we generate the co-classification matrices for each group to analyze the detailed difference in hierarchy between sexes \citep{bassett2015learning}. Females' and males' brain networks display different interaction patterns between VIN and DAN. 
Our findings provide new evidence on the sex difference of the brain network, shedding lights on future studies of brain function, cognitivity and disorder.

\section{Materials and Methods}

\subsection{Subjects and Acquisition}

We utilize resting-state fMRI data from the Southwest University Longitudinal Imaging Multimodal (SLIM) project, which has been approved by the Institutional Review Board of the Brain Imaging Center at Southwest University, Chongqing, China. 
The resting-state functional MRI images are collected using a Siemens Trio 3.0T scanner (Siemens Medical, Erlangen, Germany) at the Southwest University China Center for Brain Imaging.
They are acquired using a single-shot, gradient-recalled echo planar imaging sequence (repetition time $= 2000$ ms, echo time $= 30$ ms, flip angle $= 90^\circ$, field of view (FOV) $= 220 \times 220$ mm, thickness/slice gap $= 3/1$ mm, and voxel size $= 3.4 \times 3.4 \times 3$ mm$^3$).
All subjects are instructed to rest with closed eyes and not think of anything particular.
For a detailed description of the subject information and data acquisition parameters, please see  \cite{liu2017longitudinal}. 
Subjects are excluded if they: (a) are $< $18 years of age; (b) are $<$ 200 timepoints of acquisition time; (c) lack of sex information; (d) have been psychiatric drugs with psychiatric or neurological disorders; (e) have a history of head trauma. The final dataset comprises $N = 541$ subjects (299 females, mean age: $19.76\pm0.03$; 241 males, mean age: $19.75\pm0.03$; range: 18-27 years). 

\subsection{Construction of the functional brain network}

\begin{figure}[H]
	\begin{adjustwidth}{-\extralength}{0cm}
		\centering
		\includegraphics[width=15.5cm]{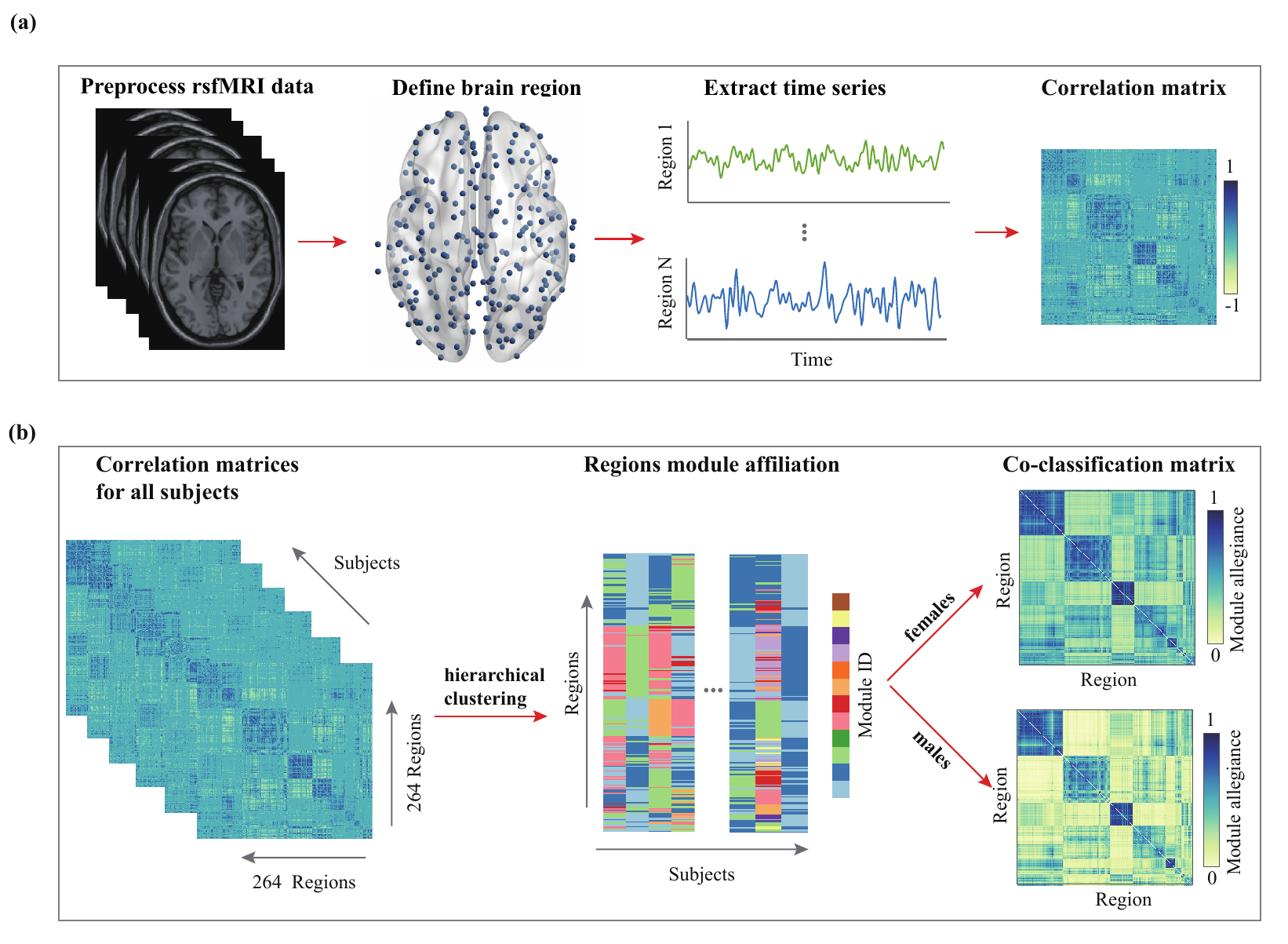}
	\end{adjustwidth}
	\caption{Procedure for Constructing a Functional Brain Network and Co-classification Matrix. \textbf{(a)} Construction of the Functional Brain Network. Standard preprocessing steps, such as motion correction, normalization, and spatial smoothing, are applied to the resting-state fMRI data to ensure high-quality input for subsequent analyses. The cerebral cortex is divided into 264 brain regions based on the Power template \citep{power2011functional}. Each region serves as a node in the brain network. The mean time series of each region is extracted.
		Pearson’s correlation coefficients are calculated for each pair of nodes, resulting in a $264 \times 264$ correlation matrix that represents the functional brain network. 
		\textbf{(b)} Construction of Co-classification Matrix. In the first stage, the modular structure for each individual's brain network is detected using hierarchical clustering and maximum modularity. In the second stage, a co-classification matrix for each group (i.e., females and males) is constructed by combining all the individual modular structures. }
	\label{fig}
\end{figure}

An overview of the processing of the construction of functional brain network for each individual is displayed in Figure \ref{fig}a, with each step explained below.

\textbf{Pre-processing rsfMRI Data}

Functional images are preprocessed using the DPARSF \citep{yan2016dpabi} and SPM8 toolits \\ (\href{https://www.fil.ion.ucl.ac.uk/spm/}{https://www.fil.ion.ucl.ac.uk/spm/}). After discarding the first 10 rs-fMRI images to ensure signal stabilization, the rest of the rs-fMRI images are corrected for slice time difference and head motion. Next, we register the images and spatially normalize them to the Montreal Neurological Institute (MNI) template, resample the voxel size to $= 3 \times 3 \times 3\,\text{mm}^3$. We then perform spatial smoothing with a $6\,\text{mm}$ full-width at half maximum Gaussian kernel (FWHM $= 6\,\text{mm}$). We clear confounding signals, including white matter, cerebrospinal fluid signals, and head motion effects, using the Friston 24-parameter model  \citep{power2012spurious,yan2013comprehensive}. Finally, we apply a temporal band-pass filter ($0.01\, \text{Hz} - 0.1\, \text{Hz}$) to reduce high-frequency physiological noise.

\textbf{Constructing network}

Nodes in functional brain networks are defined by the Power atlas \citep{power2011functional}, which includes 264 brain regions.
This atlas also offers a parcellation scheme for these nodes, delineating 13 functional modules corresponding to known large-scale brain networks coherent during both task activity and rest.
The 13 functional modules include auditory network (AUN), visual network (VIN), subcortical network (SUB), salience network (SN), cerebellar (CER), default mode network (DMN), ventral attention network (VAN), dorsal attention network (DAN), cingulo-opercular task control (COTC), fronto-parietal task control (FPTC), sensory/somatomotor hand (SMH), and sensory/somatomotor mouth (SMM).  
For each subject, the brain network is quantified by a ${264 \times 264}$ matrix $\textbf{A}$, which is calculated using Pearson correlation of averaged Blood Oxygenation Level Dependent (BOLD) signals (i.e., averaged time series) of paired nodes. 
For this brain matrix $\textbf{A}$, each element $a_{i,j}$ represents the connection strength between node $i$ and $j$, calculated from the Fisher's r-to-z normalized value. 
To ensure consensus within the same sex, we apply the proportional threshold method \citep{de2013estimating}: a connection is kept if it appears within the same sex greater than 50\% of cases. 

\subsection{Hierarchical clustering}

Hierarchical clustering is a basic and extensively utilized approach for delineating the multiscale modular hierarchy in brain network \citep{zhou2006hierarchical,vidaurre2017brain,boly2012hierarchical,akiki2019determining}. This technique facilitates the exploration of how entities are organized into modules, thereby illuminating the complexity hierarchical structures within these networks \citep{brigatti2021entropy}.
In this study, we adopt a hierarchical agglomerative clustering method, which is particularly effective in capturing the interactions among brain regions.
Given the fact that both positive and negative weights exist in the brain network connections, we first transform the connection matrix $\textbf{A}$ into a distance matrix $\textbf{D}$ by assigning the element value $d_{i,j} = 1 - a_{i,j}$. Then we perform the standard hierarchical clustering on $\textbf{D}$ to get a dendrogram. At each iteration, two entities (nodes or clusters) with the smallest distance are merged together to form a new cluster, adding a new layer in the dendrogram. The distance between two entities are calculated as the average distance of the nodes in them. The process is repeated until all nodes are merged into a single cluster.

\subsection{Network topology analysis}
The complexity and connectivity within brain networks can be explored through various metrics, each highlighting different aspects of network dynamics \citep{shumbayawonda2019sex}. In the following, we briefly introduce the metrics applied in this study. We also introduce the tool to get the aggregated module, in order to find the detailed brain region differences between females and males.

\textbf{Hierarchical entropy}

Entropy is widely used as a measure of randomness. To quantify the complexity of a hierarchical organization, several metrics are proposed such as thermodynamic entropy, Kolmogorov entropy, and hierarchical entropy \citep{chappell2014defining, jiang2011hierarchical,zhu2014roller,wang2024optimization,blair2024dynamic,kringelbach2024thermodynamics}. 

Hierarchical entropy \citep{jiang2011hierarchical,zhu2014roller} is a metric that focuses on the process of hierarchical clustering.
For the $i$th layer of the dendrogram, we can quantify the uniformity of the cluster size using Shannon-entropy as
\begin{equation}
    h_i = -\sum_{j \in J}P_{ij}\log_{2} P_{ij},
\end{equation}
where $J$ is the set of clusters on the $i$th layer, $n_{ij}$ is the number of nodes of cluster $j$ on the $i$th layer, $P_{ij} = \frac{n_{ij}}{N}$ is the size distribution of the cluster and $N$ is the total number of nodes. $h_i$ is high if multiple small clusters with similar sizes are formed, and low if a few big clusters dominate. Hence, the value of $h_i$ reflects the formation of the hierarchical structure.
By averaging $h_i$ of each layer, we have the hierarchical entropy $H$ of a hierarchical organization 
\begin{equation}
    H = \sum_{i}^{N-1} \frac{h_i}{N-1}.
\end{equation}

\textbf{Modularity}

Modularity is a metric to assess the extent to which the modules or communities are separated \citep{newman2006modularity,rubinov2011weight}. 
A higher modularity suggests that modules are further separated such that connections within each module are stronger than that in between.
The maximum modularity of a network usually refers to the best partition of nodes. In this work, we consider the maximum modularity $Q^*$ as a measure of the network partition
\begin{equation}
\label{modularity}
     Q^* =  max \{\frac{1}{2W^+}\sum_{ij}(w^+_{ij} - e^+_{ij})\delta(m_i,m_j) - \frac{1}{2W^+ + 2W^{-}}\sum_{ij}(w^-_{ij} - e^-_{ij})\delta(m_i,m_j)\},
\end{equation}
where $W$ is the total weight of all connections, $w_{ij}$ is the weighted and signed connection between region $i$ and $j$, $e_{ij}$ is the strength of a connection divided by the total weight of the network, and $\delta(m_i,m_j)$ is set to 1 when region $i$ and $j$ are in the same modular and 0 otherwise. "$+$" and "$-$" superscripts denote all positive and negative connections, respectively. 

\subsection{Co-classiffication matrix}

The co-classification matrix  $\textbf{C}$, also known as the module allegiance matrix \citep{bassett2015learning}, is the tool to merge different partition structures and reach a consensus modular structure \citep{fornito2012competitive, rubinov2011neurobiologically}. 
More specifically, we find the module structure of the brain network for each individual by performing hierarchical clustering and cutting the dendrogram at the maximum modularity. 
For the group of females and males, we calculate $c_{ij}$ which is the probability that node $i$ and $j$ were assigned to the same functional module over all subjects, which eventually yield the co-classification matrix $\textbf{C}$ (Figure \ref{fig}b). 
The element values of the matrix effectively reveals the inherent overlaps in the brain network architecture across individuals of the same sex. 
Nodes that often appear in the same module will have a high connection strength in $\textbf{C}$, whereas nodes that are rarely in one module will have a low connection strength.
Therefore, when performing hierarchical clustering on $\textbf{C}$, the modular structure obtained reflects a consensus partition that is ``averaged'' over multiple individuals \citep{chen2021rank,fang2023comprehensive,lin2023scalable}. As an example, if nodes 1, 2 and 3 are consistently assigned to the same module, we will have $c_{12} = c_{21} =c_{13}=c_{31}=c_{23} = c_{32} = 1$ and other elements equals 0. The hierarchical clustering of $\textbf{C}$ will yield a module that contains nodes 1, 2 and 3, perfectly reflects the modular structure in the population.   

\subsection{Statistical tests}

A two-sample t-test is used to compare mean differences, with significance determined at $p < 0.05$.
In addition, effect size (i.e., Cohen's $d$ value) is used to assess the magnitude of mean differences between sexes.
A positive $d$ value indicates a stronger effect in females, while a negative $d$ value indicates a stronger effect in males.

\section{Results}
\subsection{The relationship between $H$ and $Q^*$}
Several works analyze the association between the hierarchical structure and modularity of the brain network. 
Some studies report a positive correlation between them \citep{sporns2016modular,betzel2017multi}. But other studies suggest that modularity may be reduced due to dense inter-layer connections, potentially giving rise to a negative correlation \citep{meunier2010modular}. Here, we employ a linear regression to quantify the relationship between $H$ and $Q^*$, in order to provide more evidence on the relationship of the brain network's hierarchy and modularity.

\begin{figure}[H]
\includegraphics[width=13.5 cm]{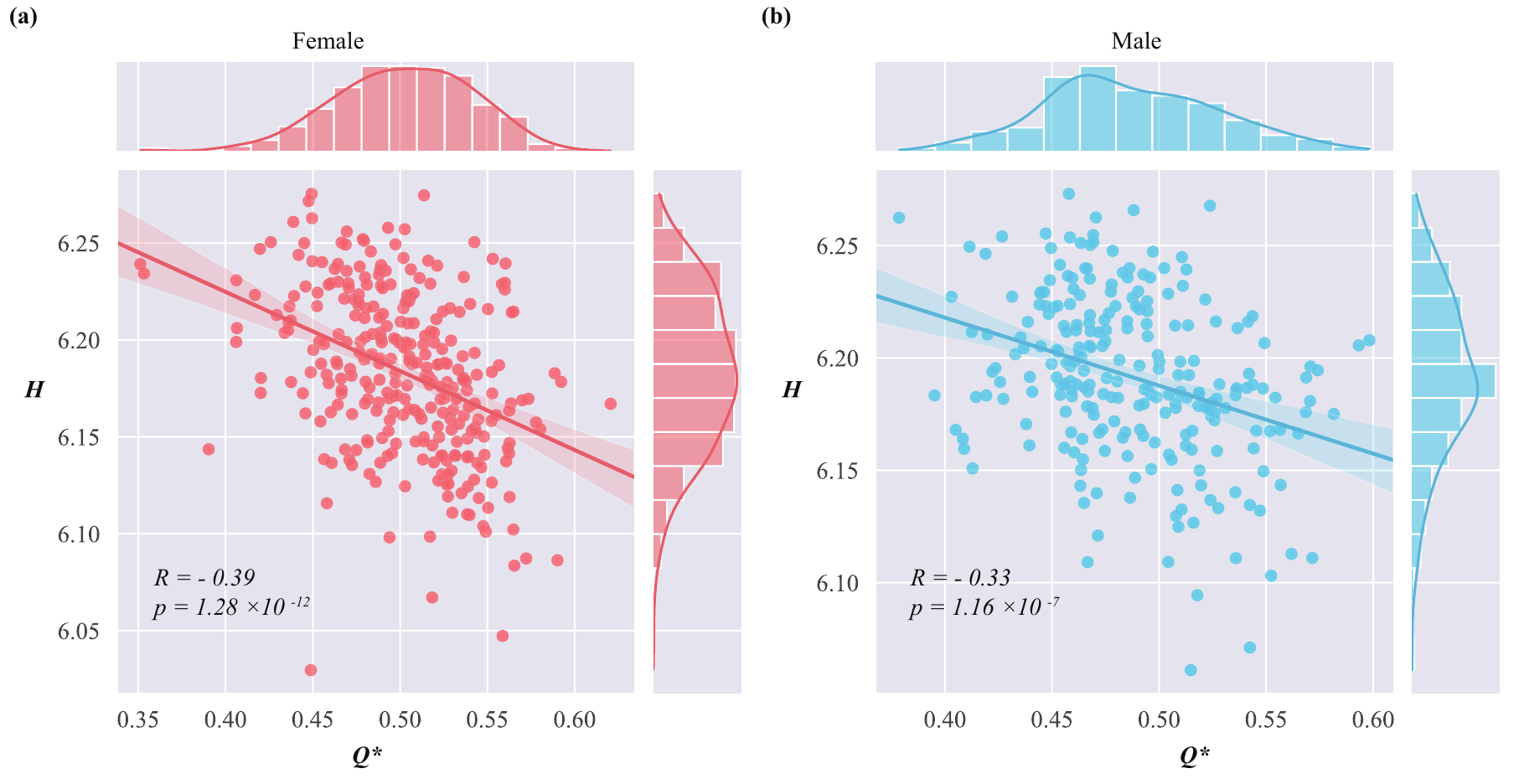}
\caption{The relationship between $H$ and $Q^*$. 
	The correlation coefficient $R$ is calculated between  $H$ and $Q^*$ , and a linear regression is performed, represented by a solid line. (\textbf{a}) Females. (\textbf{b}) Males.\label{fig1}}
\end{figure}

Figure \ref{fig1} depicts a significant negative correlation between $H$ and $Q^*$ in brain networks. 
When the complexity of the hierarchical structure becomes high, the modular structure gets diminished.
On the contrary, when the hierarchical structure becomes simpler, the modular structure is more distinct. This inverse relationship suggests a competitive dynamic between hierarchical complexity and modular clarity, potentially impacting both their functional efficiency and robustness.

\unskip

\subsection{Comparison between the sexes at the group-level}

\begin{figure}[H]
	\includegraphics[width=13.5 cm]{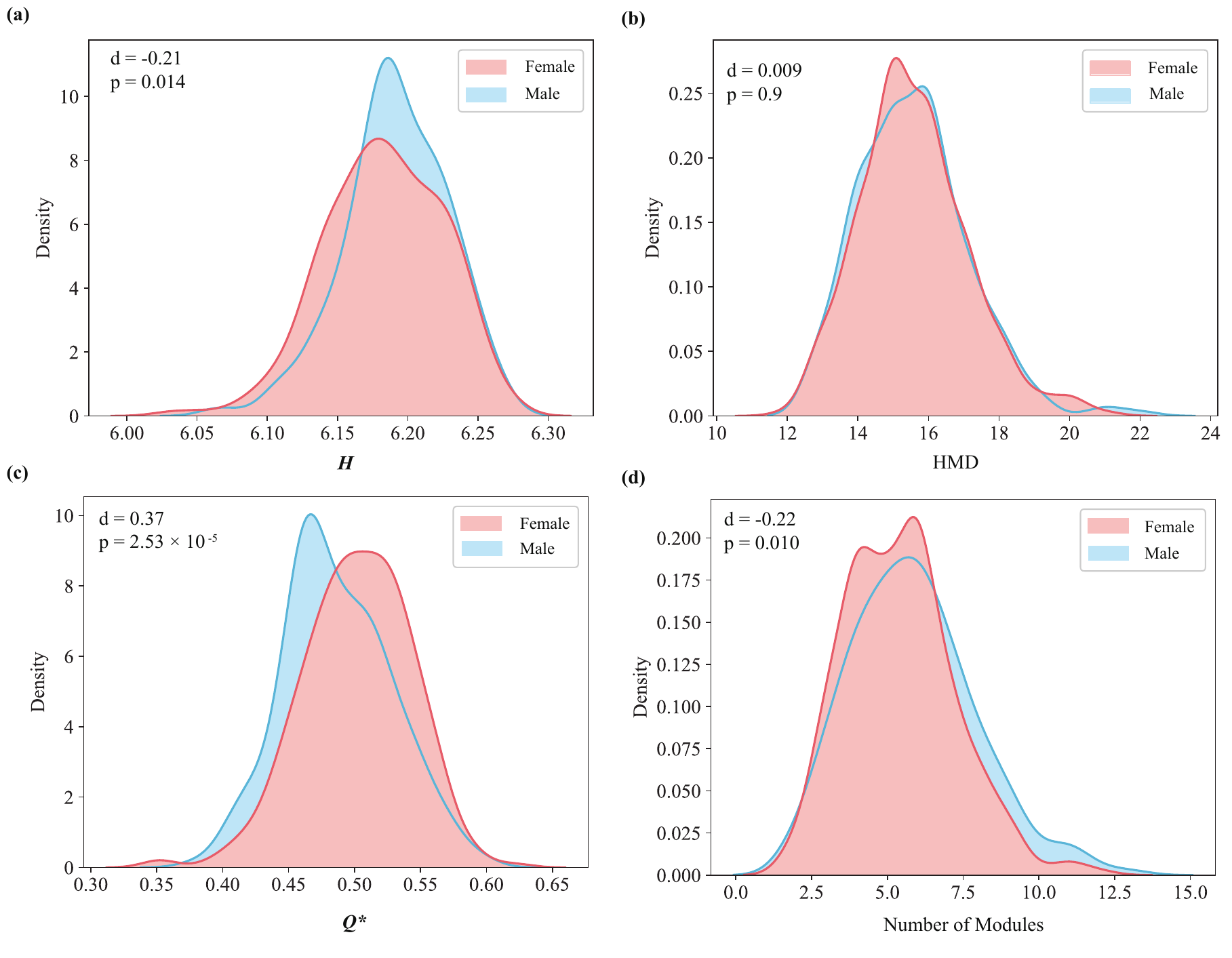}
	\caption{The distributions of topological metrics in males and females. For all figures, $d$ is the effect size and $p$ is the p-value of the two-sample t-test. (\textbf{a}) The kernel density of $H$ for each sex. (\textbf{b}) The kernel density of the maximum depth of hierarchical organization for each sex. (\textbf{c}) The kernel density of $Q^*$ for each sex. (\textbf{d}) The kernel density of the number of modules.\label{fig2}}
\end{figure}

We first compare the mean of $H$ between sexes and identify a significant difference (Figure \ref{fig2}a). The brain network of males exhibits more complex hierarchical structures than that of females. To further investigate the reasons of the difference, we compute the hierarchical maximum depth (HMD) \citep{wills1998interactive}, defined as the maximum distance from the root node to a leaf node in the dendrogram. If the dendrogram is formed such that each node iteratively merges to a single core cluster, the HMD will be very high. On the contrary, if small clusters pair with each other at each layer, the HMD will be low. We find no significant difference in HMD between sexes  (Figure \ref{fig2}b). Therefore, the different complexity in hierarchical structures is not originated from different heights of the dendrogram. The uniformity of cluster sizes at each layer, related with how the modules are formed, plays a more important role. 

To further understand the differences in hierarchical structures, we identify the maximum modularity $Q^*$ and cut the dendrogram according to $Q^*$ to get modules of the network. We find that females display a significantly higher $Q^*$ (Figure \ref{fig2}c), indicating a stronger intra-modular connectivity than males. Males display a greater number of modules (Figure \ref{fig2}d), suggesting a broader, more dispersed network organization.  

Overall, these findings imply that males have a more dispersed network organization, while females have a more integrated and coordinated network.

\unskip

\subsection{Comparison between the sexes at the consensus level}

\begin{figure}[h]
\includegraphics[width=13.5 cm]{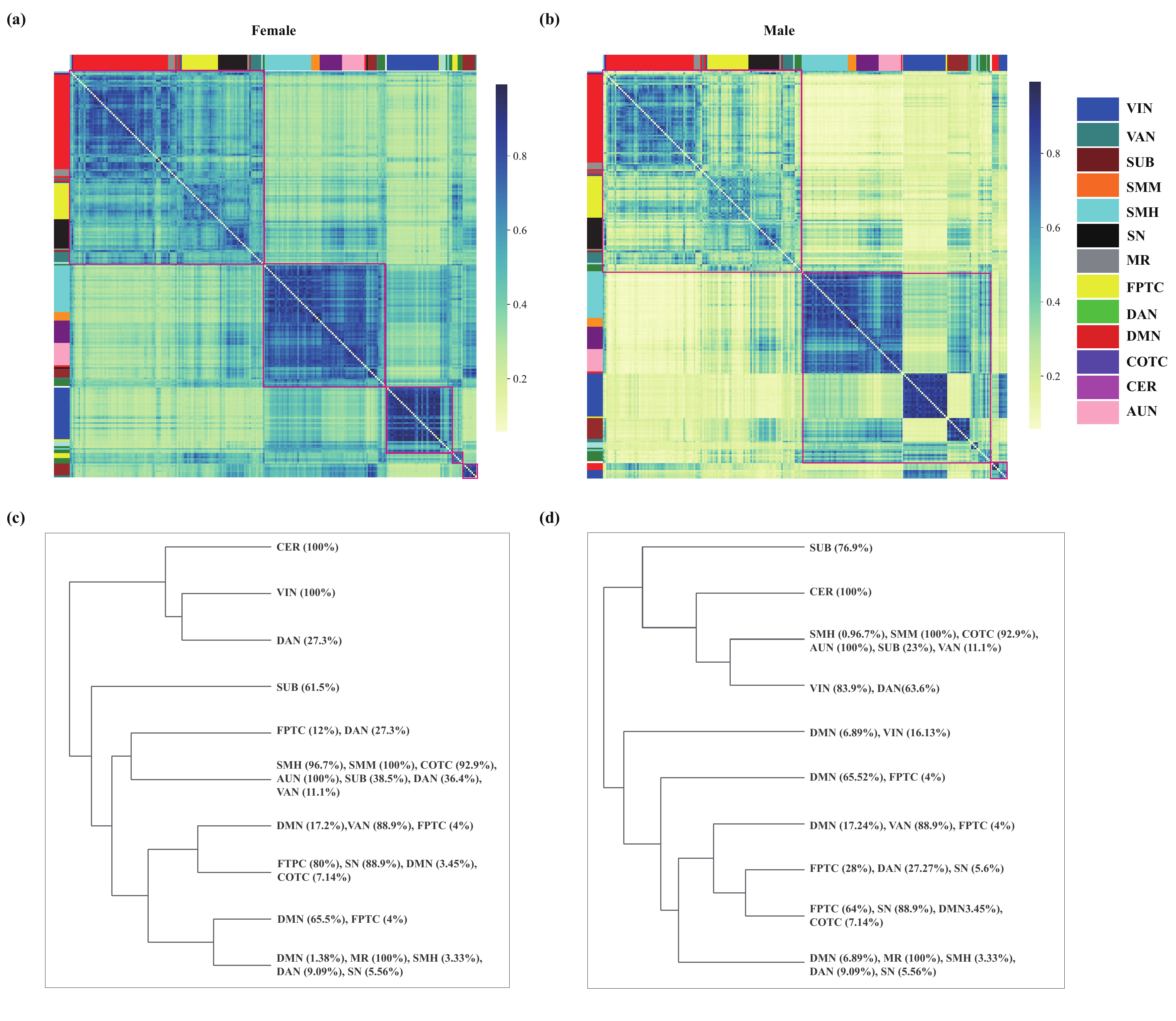}
\caption{Co-classification matrix and its hierarchical organization. (\textbf{a})-(\textbf{b}) Co-classification matrix for each sex. Red boxes represent the modular partition. The color reflect the value of the matrix element, which suggests that modular structure is more stable in females than in males. (\textbf{c})-(\textbf{d}) We extract the module compositions from the co-classification matrix for females and males respectively, and show the top-10 modules. The percentages represent the proportion of nodes from each functional network within these modules. For example, VIN (100\%) signifies that all brain regions associated with VIN are included in the same module.\label{fig3}}
\end{figure}   

To assess the detailed modular difference between females and males, we construct the co-classification matrix and apply the hierarchical clustering to get the consensus partition for each sex (Figure \ref{fig3}a-b). Overall, the element value of the co-classification matrix is higher for females than males. This suggests that the modular structure of females is more stable than that in males. For the consensus hierarchical structure, females exhibit high inter-module co-occurrence, blurring the boundaries between modules and resulting in a low modularity value ($Q^* = 0.1320$). 
But this high co-occurrence probability is also pronounced within smaller modules, eventually giving rise to 5 modules at the optimal partition.
In contrast, males show high intra-module co-occurrence and low inter-module co-occurrence, which results in a higher modularity value than that in femals ($Q^* = 0.2395$). 
Due to the comparable co-occurrence probabilities between smaller modules like SUB and VIN, and larger ones formed by SMH, SMM, COTC, and AUN, they are grouped into one large module. We identify 3 modules for males. 

We further look into the dendrogram and analyze the formation of the modules in each sex (Figure \ref{fig3}c-d). We identify distinct interactions within the DAN and VIN brain regions. In females, the DAN region exhibit frequent collaboration both within and between modules, whereas the VIN regions engage more with other areas within the same VIN module.
In males, a smaller subset of the DAN regions collaborates with specific parts of the FPTC regions, while a more substantial portion integrates extensively with the VIN regions.
These observation suggest that females and males use different integration strategies to form hierarchical organization.
In addition, we also observe some differences in the modular arrangements between the sexes. For instance, in females, modules including DMN, FPTC, SN, MR, and VAN show a higher co-occurrence probability compared to males, and areas within the DMN and VIN typically stay within their respective modules. In contrast, in males, a few nodes from DMN and VIN form an independent module.  SUB regions in females tend to interact internally, in males, they are more likely to interact with other functional modules. Overall, the modular structure in females is more stable, exhibiting higher co-occurrence probabilities and less variability than in males.

\section{Discussion}
Utilizing the resting-state fMRI dataset, we investigate whether there are sex differences in the hierarchical organization and modular structure of brain networks, and further look into their specific differences. We employ hierarchical entropy ($H$) to quantify the complexity of the brain’s hierarchical organization and maximum modularity $Q^*$ to assess the network’s modularity. Our analysis indicates a negative correlation between $H$ and $Q^*$. Specifically, female brain networks demonstrate higher $Q^*$, suggesting stronger intra-module connections, whereas male brain networks exhibit greater $H$, indicating more complex interactions among different clusters. Further, a detailed examination of the co-classification matrix, we find higher interaction probabilities within female functional modules and more frequent between functional module interactions in males, particular in DAN and VIN. These differences result in inconsistent modular structures between males and females, suggesting that they employ different organizational and integration strategies.

Our study quantitatively confirms an inverse correlation between the complexity of hierarchy and modularity in brain networks, aligning with prior findings \citep{meunier2010modular}. More complex hierarchical structures enhance cross-regional communication and information integration  \citep{hilgetag2020hierarchy,betzel2017multi}, yet they might reduce the independence and local processing efficiency of modules. Conversely, while highly modular networks excel at processing local information and ensuring functional autonomy \citep{sporns2016modular}, they could impede long-distance information transfer and collaboration across modules. 
These insights emphasize the need for a balance between hierarchical complexity and modularity to optimize brain network functionality \citep{bertolero2015modular, bassett2010efficient}. This balance enhances robustness and adaptability, crucial for responding to environmental changes and boosting cognitive performance, illustrating the critical interplay between network structure and brain function \citep{fox2005human, honey2007network, meunier2009hierarchical, de2013estimating, meunier2010modular}.

At the group level, we observe that female brain networks exhibit lower hierarchical complexity and higher modularity, in contrast to male networks which show higher hierarchical complexity and lower modularity. These observations are consistent with the negative correlation between hierarchical complexity and modularity, suggesting differing organizational strategies between sexes.
This trend is further supported by the use of hierarchical clustering methods on the co-classification matrix, where we found that females tend to integrate more within functional modules, while males engage more in interactions between modules. Such patterns lead to simpler interaction patterns in females and more complex ones in males, explaining the observed differences in complexity.
Additionally, the analysis of the consensus modular structure shows higher intra-modular co-occurrence probabilities in females compared to males, which may account for the greater modularity observed at the group level. These findings is corroborated by previous studies that indicate stronger intra-modular connections in females and stronger inter-modular connections in males \citep{ritchie2018sex, satterthwaite2015linked, allen2011baseline, biswal2010toward}, supporting our conclusions on the distinct modular and hierarchical structures across sexes.
These insights collectively underscore the profound impact of sex on brain network organization, reflecting distinct information processing strategies that have broad implications for understanding cognitive and behavioral sex differences \citep{xu2020sex,logan2010investigating,marrocco2016sex}.

Specifically, within the consensus hierarchical structure, we observe notable differences in integration styles within the VIN between males and females. In female brains, the VIN regions primarily engage in integration within the same functional module, indicating that this robust intra-modular integration underlies their enhanced abilities in processing visual details, recognizing facial expressions, and perceiving emotions \citep{loven2011women,nolen2012emotion, mak2009sex}.
On the contrary male brains show a tendency for the VIN regions to integrate more extensively with the DAN and the DMN. This pattern highlights males’ strengths in tasks that require spatial orientation and environmental awareness \citep{basso2004global,cimadevilla2020spatial}. Such inter-modular integration is crucial for combining complex environmental information, which supports spatial navigation, multitasking, and rapid responses in urgent situations \citep{munion2019gender,mantyla2013gender,generoso2016simulation}.

However, the current study also has several limitations.
First, although we observe most sex differences based on resting-state fMRI, it does not capture task-specific brain activations \citep{lurie2020questions,cheng2018principal}, which might be crucial for understanding functional differences between male and female brains.
Second, $H$ provides a quantitative measure of complexity but does not offer insights into the functional implications of the observed differences.
Higher or lower entropy values might not directly correlate with specific cognitive or behavioral traits, making it difficult to draw definitive conclusions about sex differences.
Finally, we analyze interaction differences in modules at a specific layer (i.e., the structure with maximum modularity) but did not consider multi-level structural differences.
Given these limitations, future work should aim to combine analyses of the multi-scale hierarchical structures of resting-state functional brain networks with task-specific paradigms.
This integrative approach may facilitate a deeper understanding of sex differences in daily behaviors and cognitive functions.
\section{Conclusions}
In summary, this study conducts a systematic analysis of the relationship between hierarchical organization and modular structure of brain networks, clearly illustrating the impact of sex on these structural attributes. 
Our findings identify a pronounced negative correlation: the more complex a brain network’s hierarchical organization, the lower its modularity. Notably, female brains demonstrate enhanced modularity, while male brains display more complex hierarchical structures. 
These distinctions underscore the significance of including sex considerations in neuroscience research, particularly when investigating the mechanisms by which the brain processes and integrates information. 
At the consensus level, we have observed that male and female brains employ distinct strategies for organization and integration, which highlights specific sex differences in how information is processed and integrated.
Future research should delve deeper into how these sex-specific differences in brain network structures influence cognitive and behavioral functions. 
Furthermore, as these structural variations may influence disease susceptibility and treatment responses, further research is essential to develop optimized prevention and intervention strategies based on sex differences.

\funding{The work is supported by Chongqing Innovative Research Groups (No. CXQT21005) and the Fundamental Research Funds for the Central Universities (No. SWU-XDJH202303).}

\institutionalreview{Not applicable.}

\informedconsent{Not applicable.}

\dataavailability{Not applicable.} 

\conflictsofinterest{The authors declare no conflicts of interest.} 




\begin{adjustwidth}{-\extralength}{0cm}

\reftitle{References}


\bibliography{referen}

\begin{thebibliography}{999}

\bibitem[Beery and Zucker(2011)]{beery2011sex}
Beery, A.K.; Zucker, I.
\newblock Sex bias in neuroscience and biomedical research.
\newblock {\em Neurosci. Biobehav. Rev.} {\bf 2011}, {\em 35},~565--572.

\bibitem[Pawluski et~al.(2020)Pawluski, Kokras, Charlier, and
  Dalla]{pawluski2020sex}
Pawluski, J.L.; Kokras, N.; Charlier, T.D.; Dalla, C.
\newblock Sex matters in neuroscience and neuropsychopharmacology.
\newblock {\em Eur. J. Neurosci.} {\bf 2020}, {\em 52},~2423--2428.

\bibitem[Hawkes and Chang(2024)]{hawkes2024sex}
Hawkes, S.J.; Chang, A.Y.
\newblock Time to implement sex and gender responsive policies and programmes.
\newblock {\em Lancet Public Health} {\bf 2024}, {\em 9},~e276--e277.

\bibitem[Nolen-Hoeksema(2012)]{nolen2012emotion}
Nolen-Hoeksema, S.
\newblock Emotion regulation and psychopathology: The role of gender.
\newblock {\em Annu. Rev. Clin. Psycho.} {\bf 2012}, {\em 8},~161--187.

\bibitem[Mak et~al.(2009)Mak, Hu, Zhang, Xiao, and Lee]{mak2009sex}
Mak, A.K.; Hu, Z.g.; Zhang, J.X.; Xiao, Z.; Lee, T.M.
\newblock Sex-related differences in neural activity during emotion regulation.
\newblock {\em Neuropsychologia} {\bf 2009}, {\em 47},~2900--2908.

\bibitem[Campbell-Sills et~al.(2006)Campbell-Sills, Barlow, Brown, and
  Hofmann]{campbell2006acceptability}
Campbell-Sills, L.; Barlow, D.H.; Brown, T.A.; Hofmann, S.G.
\newblock Acceptability and suppression of negative emotion in anxiety and mood
  disorders.
\newblock {\em Emotion} {\bf 2006}, {\em 6},~587.

\bibitem[Garnefski et~al.(2004)Garnefski, Teerds, Kraaij, Legerstee, and van
  Den~Kommer]{garnefski2004cognitive}
Garnefski, N.; Teerds, J.; Kraaij, V.; Legerstee, J.; van Den~Kommer, T.
\newblock Cognitive emotion regulation strategies and depressive symptoms:
  Differences between males and females.
\newblock {\em Pers. Indiv. Differ.} {\bf 2004}, {\em 36},~267--276.

\bibitem[Joormann et~al.(2010)Joormann, Gilbert, and
  Gotlib]{joormann2010emotion}
Joormann, J.; Gilbert, K.; Gotlib, I.H.
\newblock Emotion identification in girls at high risk for depression.
\newblock {\em J. Child. Psychol. Psyc.} {\bf 2010}, {\em 51},~575--582.

\bibitem[Deckert et~al.(2020)Deckert, Schmoeger, Auff, and
  Willinger]{deckert2020subjective}
Deckert, M.; Schmoeger, M.; Auff, E.; Willinger, U.
\newblock Subjective emotional arousal: an explorative study on the role of
  gender, age, intensity, emotion regulation difficulties, depression and
  anxiety symptoms, and meta-emotion.
\newblock {\em Psychol. Res.} {\bf 2020}, {\em 84},~1857--1876.

\bibitem[Luo et~al.(2022)Luo, Chen, Qiu, and Jia]{luo2022accelerated}
Luo, Y.; Chen, W.; Qiu, J.; Jia, T.
\newblock Accelerated functional brain aging in major depressive disorder:
  evidence from a large scale fMRI analysis of Chinese participants.
\newblock {\em Transl. Psychiatry} {\bf 2022}, {\em 12},~397.

\bibitem[Cimadevilla and Piccardi(2020)]{cimadevilla2020spatial}
Cimadevilla, J.M.; Piccardi, L.
\newblock Spatial skills.
\newblock {\em Handbook of clinical neurology} {\bf 2020}, {\em 175},~65--79.

\bibitem[Lawrence et~al.(2020)Lawrence, Hernandez, Bowman, Padgaonkar, Fuster,
  Jack, Aylward, Gaab, Van~Horn, Bernier, et~al.]{lawrence2020sex}
Lawrence, K.E.; Hernandez, L.M.; Bowman, H.C.; Padgaonkar, N.T.; Fuster, E.;
  Jack, A.; Aylward, E.; Gaab, N.; Van~Horn, J.D.; Bernier, R.A.;  et~al.
\newblock Sex differences in functional connectivity of the salience, default
  mode, and central executive networks in youth with ASD.
\newblock {\em Cereb. Cortex} {\bf 2020}, {\em 30},~5107--5120.

\bibitem[Rosch et~al.(2018)Rosch, Mostofsky, and Nebel]{rosch2018adhd}
Rosch, K.S.; Mostofsky, S.H.; Nebel, M.B.
\newblock ADHD-related sex differences in fronto-subcortical intrinsic
  functional connectivity and associations with delay discounting.
\newblock {\em J. Neurodev. Disord.} {\bf 2018}, {\em 10},~1--14.

\bibitem[Logothetis et~al.(2001)Logothetis, Pauls, Augath, Trinath, and
  Oeltermann]{logothetis2001neurophysiological}
Logothetis, N.K.; Pauls, J.; Augath, M.; Trinath, T.; Oeltermann, A.
\newblock Neurophysiological investigation of the basis of the fMRI signal.
\newblock {\em Nature} {\bf 2001}, {\em 412},~150--157.

\bibitem[Bullmore and Sporns(2009)]{bullmore2009complex}
Bullmore, E.; Sporns, O.
\newblock Complex brain networks: graph theoretical analysis of structural and
  functional systems.
\newblock {\em Nat. Rev. Neurosci.} {\bf 2009}, {\em 10},~186--198.

\bibitem[Bassett and Sporns(2017)]{bassett2017network}
Bassett, D.S.; Sporns, O.
\newblock Network neuroscience.
\newblock {\em Nat. Neurosci.} {\bf 2017}, {\em 20},~353--364.

\bibitem[Satterthwaite et~al.(2015)Satterthwaite, Wolf, Roalf, Ruparel, Erus,
  Vandekar, Gennatas, Elliott, Smith, Hakonarson,
  et~al.]{satterthwaite2015linked}
Satterthwaite, T.D.; Wolf, D.H.; Roalf, D.R.; Ruparel, K.; Erus, G.; Vandekar,
  S.; Gennatas, E.D.; Elliott, M.A.; Smith, A.; Hakonarson, H.;  et~al.
\newblock Linked sex differences in cognition and functional connectivity in
  youth.
\newblock {\em Cereb. Cortex} {\bf 2015}, {\em 25},~2383--2394.

\bibitem[Zhang et~al.(2018)Zhang, Dougherty, Baum, White, and
  Michael]{zhang2018functional}
Zhang, C.; Dougherty, C.C.; Baum, S.A.; White, T.; Michael, A.M.
\newblock Functional connectivity predicts gender: Evidence for gender
  differences in resting brain connectivity.
\newblock {\em Hum. Brain Mapp.} {\bf 2018}, {\em 39},~1765--1776.

\bibitem[Ritchie et~al.(2018)Ritchie, Cox, Shen, Lombardo, Reus, Alloza,
  Harris, Alderson, Hunter, Neilson, et~al.]{ritchie2018sex}
Ritchie, S.J.; Cox, S.R.; Shen, X.; Lombardo, M.V.; Reus, L.M.; Alloza, C.;
  Harris, M.A.; Alderson, H.L.; Hunter, S.; Neilson, E.;  et~al.
\newblock Sex differences in the adult human brain: evidence from 5216 UK
  biobank participants.
\newblock {\em Cereb. Cortex} {\bf 2018}, {\em 28},~2959--2975.

\bibitem[Bullmore and Sporns(2012)]{bullmore2012economy}
Bullmore, E.; Sporns, O.
\newblock The economy of brain network organization.
\newblock {\em Nat. Rev. Neurosci.} {\bf 2012}, {\em 13},~336--349.

\bibitem[Barab{\'a}si(2016)]{barabasi2016network}
Barab{\'a}si, A.L.
\newblock {\em Network Science}; Cambridge University Press,  2016.

\bibitem[Bassett et~al.(2010)Bassett, Greenfield, Meyer-Lindenberg, Weinberger,
  Moore, and Bullmore]{bassett2010efficient}
Bassett, D.S.; Greenfield, D.L.; Meyer-Lindenberg, A.; Weinberger, D.R.; Moore,
  S.W.; Bullmore, E.T.
\newblock Efficient physical embedding of topologically complex information
  processing networks in brains and computer circuits.
\newblock {\em PLoS Comp. Biol.} {\bf 2010}, {\em 6},~e1000748.

\bibitem[Sporns and Betzel(2016)]{sporns2016modular}
Sporns, O.; Betzel, R.F.
\newblock Modular brain networks.
\newblock {\em Annu. Rev. Psychol.} {\bf 2016}, {\em 67},~613--640.

\bibitem[Betzel and Bassett(2017)]{betzel2017multi}
Betzel, R.F.; Bassett, D.S.
\newblock Multi-scale brain networks.
\newblock {\em NeuroImage} {\bf 2017}, {\em 160},~73--83.

\bibitem[Betzel et~al.(2014)Betzel, Byrge, He, Go{\~n}i, Zuo, and
  Sporns]{betzel2014changes}
Betzel, R.F.; Byrge, L.; He, Y.; Go{\~n}i, J.; Zuo, X.N.; Sporns, O.
\newblock Changes in structural and functional connectivity among resting-state
  networks across the human lifespan.
\newblock {\em NeuroImage} {\bf 2014}, {\em 102},~345--357.

\bibitem[Betzel et~al.(2015)Betzel, Mi{\v{s}}i{\'c}, He, Rumschlag, Zuo, and
  Sporns]{betzel2015functional}
Betzel, R.F.; Mi{\v{s}}i{\'c}, B.; He, Y.; Rumschlag, J.; Zuo, X.N.; Sporns, O.
\newblock Functional brain modules reconfigure at multiple scales across the
  human lifespan.
\newblock {\em arXiv preprint arXiv:1510.08045} {\bf 2015}.

\bibitem[Meunier et~al.(2010)Meunier, Lambiotte, and
  Bullmore]{meunier2010modular}
Meunier, D.; Lambiotte, R.; Bullmore, E.T.
\newblock Modular and hierarchically modular organization of brain networks.
\newblock {\em Front. Neurosci.} {\bf 2010}, {\em 4},~200.

\bibitem[Meunier et~al.(2009)Meunier, Lambiotte, Fornito, Ersche, and
  Bullmore]{meunier2009hierarchical}
Meunier, D.; Lambiotte, R.; Fornito, A.; Ersche, K.; Bullmore, E.T.
\newblock Hierarchical modularity in human brain functional networks.
\newblock {\em Front. Neuroinform.} {\bf 2009}, {\em 3},~571.

\bibitem[Esfahlani et~al.(2021)Esfahlani, Jo, Puxeddu, Merritt, Tanner,
  Greenwell, Patel, Faskowitz, and Betzel]{esfahlani2021modularity}
Esfahlani, F.Z.; Jo, Y.; Puxeddu, M.G.; Merritt, H.; Tanner, J.C.; Greenwell,
  S.; Patel, R.; Faskowitz, J.; Betzel, R.F.
\newblock Modularity maximization as a flexible and generic framework for brain
  network exploratory analysis.
\newblock {\em NeuroImage} {\bf 2021}, {\em 244},~118607.

\bibitem[Ingalhalikar et~al.(2014)Ingalhalikar, Smith, Parker, Satterthwaite,
  Elliott, Ruparel, Hakonarson, Gur, Gur, and Verma]{ingalhalikar2014sex}
Ingalhalikar, M.; Smith, A.; Parker, D.; Satterthwaite, T.D.; Elliott, M.A.;
  Ruparel, K.; Hakonarson, H.; Gur, R.E.; Gur, R.C.; Verma, R.
\newblock Sex differences in the structural connectome of the human brain.
\newblock {\em PNAS} {\bf 2014}, {\em 111},~823--828.

\bibitem[Fauchon et~al.(2021)Fauchon, Meunier, Rogachov, Hemington, Cheng,
  Bosma, Osborne, Kim, Hung, Inman, et~al.]{fauchon2021sex}
Fauchon, C.; Meunier, D.; Rogachov, A.; Hemington, K.S.; Cheng, J.C.; Bosma,
  R.L.; Osborne, N.R.; Kim, J.A.; Hung, P.S.P.; Inman, R.D.;  et~al.
\newblock Sex differences in brain modular organization in chronic pain.
\newblock {\em Pain} {\bf 2021}, {\em 162},~1188--1200.

\bibitem[Stumme et~al.(2020)Stumme, Jockwitz, Hoffstaedter, Amunts, and
  Caspers]{stumme2020functional}
Stumme, J.; Jockwitz, C.; Hoffstaedter, F.; Amunts, K.; Caspers, S.
\newblock Functional network reorganization in older adults: Graph-theoretical
  analyses of age, cognition and sex.
\newblock {\em NeuroImage} {\bf 2020}, {\em 214},~116756.

\bibitem[Li et~al.(2021)Li, Dahmani, Wang, Ren, Stocklein, Lin, Luan, Zhang,
  Lu, Gali{\`e}, et~al.]{li2021co}
Li, M.; Dahmani, L.; Wang, D.; Ren, J.; Stocklein, S.; Lin, Y.; Luan, G.;
  Zhang, Z.; Lu, G.; Gali{\`e}, F.;  et~al.
\newblock Co-activation patterns across multiple tasks reveal robust
  anti-correlated functional networks.
\newblock {\em NeuroImage} {\bf 2021}, {\em 227},~117680.

\bibitem[Mitra and Raichle(2018)]{mitra2018principles}
Mitra, A.; Raichle, M.E.
\newblock Principles of cross-network communication in human resting state
  fMRI.
\newblock {\em Scand. J. Psychol.} {\bf 2018}, {\em 59},~83--90.

\bibitem[Baldassarre et~al.(2014)Baldassarre, Ramsey, Hacker, Callejas,
  Astafiev, Metcalf, Zinn, Rengachary, Snyder, Carter,
  et~al.]{baldassarre2014large}
Baldassarre, A.; Ramsey, L.; Hacker, C.L.; Callejas, A.; Astafiev, S.V.;
  Metcalf, N.V.; Zinn, K.; Rengachary, J.; Snyder, A.Z.; Carter, A.R.;  et~al.
\newblock Large-scale changes in network interactions as a physiological
  signature of spatial neglect.
\newblock {\em Brain} {\bf 2014}, {\em 137},~3267--3283.

\bibitem[Ramkiran et~al.(2019)Ramkiran, Sharma, and Rao]{ramkiran2019resting}
Ramkiran, S.; Sharma, A.; Rao, N.P.
\newblock Resting-state anticorrelated networks in Schizophrenia.
\newblock {\em Psychiatry Res. Neuroimaging} {\bf 2019}, {\em 284},~1--8.

\bibitem[Fox et~al.(2005)Fox, Snyder, Vincent, Corbetta, Van~Essen, and
  Raichle]{fox2005human}
Fox, M.D.; Snyder, A.Z.; Vincent, J.L.; Corbetta, M.; Van~Essen, D.C.; Raichle,
  M.E.
\newblock The human brain is intrinsically organized into dynamic,
  anticorrelated functional networks.
\newblock {\em PNAS} {\bf 2005}, {\em 102},~9673--9678.

\bibitem[Spreng et~al.(2016)Spreng, Stevens, Viviano, and
  Schacter]{spreng2016attenuated}
Spreng, R.N.; Stevens, W.D.; Viviano, J.D.; Schacter, D.L.
\newblock Attenuated anticorrelation between the default and dorsal attention
  networks with aging: evidence from task and rest.
\newblock {\em Neurobiol. Aging} {\bf 2016}, {\em 45},~149--160.

\bibitem[Bassett et~al.(2015)Bassett, Yang, Wymbs, and
  Grafton]{bassett2015learning}
Bassett, D.S.; Yang, M.; Wymbs, N.F.; Grafton, S.T.
\newblock Learning-induced autonomy of sensorimotor systems.
\newblock {\em Nat. Neurosci.} {\bf 2015}, {\em 18},~744--751.

\bibitem[Liu et~al.(2017)Liu, Wei, Chen, Yang, Meng, Wu, Bi, Zhang, Zuo, and
  Qiu]{liu2017longitudinal}
Liu, W.; Wei, D.; Chen, Q.; Yang, W.; Meng, J.; Wu, G.; Bi, T.; Zhang, Q.; Zuo,
  X.N.; Qiu, J.
\newblock Longitudinal test-retest neuroimaging data from healthy young adults
  in southwest China.
\newblock {\em Scientific Data} {\bf 2017}, {\em 4},~1--9.

\bibitem[Power et~al.(2011)Power, Cohen, Nelson, Wig, Barnes, Church, Vogel,
  Laumann, Miezin, Schlaggar, et~al.]{power2011functional}
Power, J.D.; Cohen, A.L.; Nelson, S.M.; Wig, G.S.; Barnes, K.A.; Church, J.A.;
  Vogel, A.C.; Laumann, T.O.; Miezin, F.M.; Schlaggar, B.L.;  et~al.
\newblock Functional network organization of the human brain.
\newblock {\em Neuron} {\bf 2011}, {\em 72},~665--678.

\bibitem[Yan et~al.(2016)Yan, Wang, Zuo, and Zang]{yan2016dpabi}
Yan, C.G.; Wang, X.D.; Zuo, X.N.; Zang, Y.F.
\newblock DPABI: data processing \& analysis for (resting-state) brain imaging.
\newblock {\em Neuroinformatics} {\bf 2016}, {\em 14},~339--351.

\bibitem[Power et~al.(2012)Power, Barnes, Snyder, Schlaggar, and
  Petersen]{power2012spurious}
Power, J.D.; Barnes, K.A.; Snyder, A.Z.; Schlaggar, B.L.; Petersen, S.E.
\newblock Spurious but systematic correlations in functional connectivity MRI
  networks arise from subject motion.
\newblock {\em NeuroImage} {\bf 2012}, {\em 59},~2142--2154.

\bibitem[Yan et~al.(2013)Yan, Cheung, Kelly, Colcombe, Craddock, Di~Martino,
  Li, Zuo, Castellanos, and Milham]{yan2013comprehensive}
Yan, C.G.; Cheung, B.; Kelly, C.; Colcombe, S.; Craddock, R.C.; Di~Martino, A.;
  Li, Q.; Zuo, X.N.; Castellanos, F.X.; Milham, M.P.
\newblock A comprehensive assessment of regional variation in the impact of
  head micromovements on functional connectomics.
\newblock {\em NeuroImage} {\bf 2013}, {\em 76},~183--201.

\bibitem[de~Reus and van~den Heuvel(2013)]{de2013estimating}
de~Reus, M.A.; van~den Heuvel, M.P.
\newblock Estimating false positives and negatives in brain networks.
\newblock {\em NeuroImage} {\bf 2013}, {\em 70},~402--409.

\bibitem[Zhou et~al.(2006)Zhou, Zemanov{\'a}, Zamora, Hilgetag, and
  Kurths]{zhou2006hierarchical}
Zhou, C.; Zemanov{\'a}, L.; Zamora, G.; Hilgetag, C.C.; Kurths, J.
\newblock Hierarchical organization unveiled by functional connectivity in
  complex brain networks.
\newblock {\em Phys. Rev. Lett.} {\bf 2006}, {\em 97},~238103.

\bibitem[Vidaurre et~al.(2017)Vidaurre, Smith, and Woolrich]{vidaurre2017brain}
Vidaurre, D.; Smith, S.M.; Woolrich, M.W.
\newblock Brain network dynamics are hierarchically organized in time.
\newblock {\em PNAS} {\bf 2017}, {\em 114},~12827--12832.

\bibitem[Boly et~al.(2012)Boly, Perlbarg, Marrelec, Schabus, Laureys, Doyon,
  P{\'e}l{\'e}grini-Issac, Maquet, and Benali]{boly2012hierarchical}
Boly, M.; Perlbarg, V.; Marrelec, G.; Schabus, M.; Laureys, S.; Doyon, J.;
  P{\'e}l{\'e}grini-Issac, M.; Maquet, P.; Benali, H.
\newblock Hierarchical clustering of brain activity during human nonrapid eye
  movement sleep.
\newblock {\em PNAS} {\bf 2012}, {\em 109},~5856--5861.

\bibitem[Akiki and Abdallah(2019)]{akiki2019determining}
Akiki, T.J.; Abdallah, C.G.
\newblock Determining the hierarchical architecture of the human brain using
  subject-level clustering of functional networks.
\newblock {\em Scientific Reports} {\bf 2019}, {\em 9},~19290.

\bibitem[Brigatti et~al.(2021)Brigatti, Netto, de~Sousa~Filho, and
  Cacholas]{brigatti2021entropy}
Brigatti, E.; Netto, V.; de~Sousa~Filho, F.; Cacholas, C.
\newblock Entropy and hierarchical clustering: Characterizing the morphology of
  the urban fabric in different spatial cultures.
\newblock {\em Chaos} {\bf 2021}, {\em 31}.

\bibitem[Shumbayawonda et~al.(2019)Shumbayawonda, Ab{\'a}solo, L{\'o}pez-Sanz,
  Bru{\~n}a, Maestu, and Fern{\'a}ndez]{shumbayawonda2019sex}
Shumbayawonda, E.; Ab{\'a}solo, D.; L{\'o}pez-Sanz, D.; Bru{\~n}a, R.; Maestu,
  F.; Fern{\'a}ndez, A.
\newblock Sex differences in the complexity of healthy older adults’
  magnetoencephalograms.
\newblock {\em Entropy} {\bf 2019}, {\em 21},~798.

\bibitem[Chappell and Dewey(2014)]{chappell2014defining}
Chappell, D.; Dewey, T.G.
\newblock Defining the entropy of hierarchical organizations.
\newblock {\em Complexity, Governance \& Networks} {\bf 2014}, {\em 1},~41--56.

\bibitem[Jiang et~al.(2011)Jiang, Peng, and Xu]{jiang2011hierarchical}
Jiang, Y.; Peng, C.K.; Xu, Y.
\newblock Hierarchical entropy analysis for biological signals.
\newblock {\em J. Comput. Appl. Math.} {\bf 2011}, {\em 236},~728--742.

\bibitem[Zhu et~al.(2014)Zhu, Song, and Xue]{zhu2014roller}
Zhu, K.; Song, X.; Xue, D.
\newblock A roller bearing fault diagnosis method based on hierarchical entropy
  and support vector machine with particle swarm optimization algorithm.
\newblock {\em Measurement} {\bf 2014}, {\em 47},~669--675.

\bibitem[Wang et~al.(2024)Wang, Cheng, and Chen]{wang2024optimization}
Wang, J.; Cheng, F.; Chen, C.
\newblock Optimization and Evaluation of Tourism Mascot Design Based on
  Analytic Hierarchy Process--Entropy Weight Method.
\newblock {\em Entropy} {\bf 2024}, {\em 26},~585.

\bibitem[Blair et~al.(2024)Blair, Miller, and Calhoun]{blair2024dynamic}
Blair, D.S.; Miller, R.L.; Calhoun, V.D.
\newblock A Dynamic Entropy Approach Reveals Reduced Functional Network
  Connectivity Trajectory Complexity in Schizophrenia.
\newblock {\em Entropy} {\bf 2024}, {\em 26},~545.

\bibitem[Kringelbach et~al.(2024)Kringelbach, Perl, and
  Deco]{kringelbach2024thermodynamics}
Kringelbach, M.L.; Perl, Y.S.; Deco, G.
\newblock The Thermodynamics of Mind.
\newblock {\em Trends. Cogn. Sci.} {\bf 2024}.

\bibitem[Newman(2006)]{newman2006modularity}
Newman, M.E.
\newblock Modularity and community structure in networks.
\newblock {\em PNAS} {\bf 2006}, {\em 103},~8577--8582.

\bibitem[Rubinov and Sporns(2011)]{rubinov2011weight}
Rubinov, M.; Sporns, O.
\newblock Weight-conserving characterization of complex functional brain
  networks.
\newblock {\em NeuroImage} {\bf 2011}, {\em 56},~2068--2079.

\bibitem[Fornito et~al.(2012)Fornito, Harrison, Zalesky, and
  Simons]{fornito2012competitive}
Fornito, A.; Harrison, B.J.; Zalesky, A.; Simons, J.S.
\newblock Competitive and cooperative dynamics of large-scale brain functional
  networks supporting recollection.
\newblock {\em PNAS} {\bf 2012}, {\em 109},~12788--12793.

\bibitem[Rubinov et~al.(2011)Rubinov, Sporns, Thivierge, and
  Breakspear]{rubinov2011neurobiologically}
Rubinov, M.; Sporns, O.; Thivierge, J.P.; Breakspear, M.
\newblock Neurobiologically realistic determinants of self-organized
  criticality in networks of spiking neurons.
\newblock {\em PLoS Comp. Biol.} {\bf 2011}, {\em 7},~e1002038.

\bibitem[Chen et~al.(2021)Chen, Zhu, and Jia]{chen2021rank}
Chen, W.; Zhu, Z.; Jia, T.
\newblock The rank boost by inconsistency in university rankings: Evidence from
  14 rankings of Chinese universities.
\newblock {\em Quantitative Science Studies} {\bf 2021}, {\em 2},~335--349.

\bibitem[Fang et~al.(2023)Fang, Li, Li, Gao, Jia, and
  Zhang]{fang2023comprehensive}
Fang, U.; Li, M.; Li, J.; Gao, L.; Jia, T.; Zhang, Y.
\newblock A comprehensive survey on multi-view clustering.
\newblock {\em IEEE Trans. Knowledge Data Eng.} {\bf 2023}, {\em
  35},~12350--12368.

\bibitem[Lin et~al.(2023)Lin, Li, and Jia]{lin2023scalable}
Lin, L.; Li, R.; Jia, T.
\newblock Scalable and effective conductance-based graph clustering.
\newblock In Proceedings of the Proceedings of the AAAI Conference on
  Artificial Intelligence,  2023, Vol.~37, pp. 4471--4478.

\bibitem[wil(1998)]{wills1998interactive}
An interactive view for hierarchical clustering.
\newblock In Proceedings of the Proceedings IEEE Symposium on Information
  Visualization (Cat. No. 98TB100258). IEEE,  1998, pp. 26--31.

\bibitem[Hilgetag and Goulas(2020)]{hilgetag2020hierarchy}
Hilgetag, C.C.; Goulas, A.
\newblock ‘Hierarchy’in the organization of brain networks.
\newblock {\em Philos. Trans. R. Soc. B} {\bf 2020}, {\em 375},~20190319.

\bibitem[Bertolero et~al.(2015)Bertolero, Yeo, and
  D’Esposito]{bertolero2015modular}
Bertolero, M.A.; Yeo, B.T.; D’Esposito, M.
\newblock The modular and integrative functional architecture of the human
  brain.
\newblock {\em PNAS} {\bf 2015}, {\em 112},~E6798--E6807.

\bibitem[Honey et~al.(2007)Honey, K{\"o}tter, Breakspear, and
  Sporns]{honey2007network}
Honey, C.J.; K{\"o}tter, R.; Breakspear, M.; Sporns, O.
\newblock Network structure of cerebral cortex shapes functional connectivity
  on multiple time scales.
\newblock {\em PNAS} {\bf 2007}, {\em 104},~10240--10245.

\bibitem[Allen et~al.(2011)Allen, Erhardt, Damaraju, Gruner, Segall, Silva,
  Havlicek, Rachakonda, Fries, Kalyanam, et~al.]{allen2011baseline}
Allen, E.A.; Erhardt, E.B.; Damaraju, E.; Gruner, W.; Segall, J.M.; Silva,
  R.F.; Havlicek, M.; Rachakonda, S.; Fries, J.; Kalyanam, R.;  et~al.
\newblock A baseline for the multivariate comparison of resting-state networks.
\newblock {\em Front. Syst. Neurosci.} {\bf 2011}, {\em 5},~2.

\bibitem[Biswal et~al.(2010)Biswal, Mennes, Zuo, Gohel, Kelly, Smith, Beckmann,
  Adelstein, Buckner, Colcombe, et~al.]{biswal2010toward}
Biswal, B.B.; Mennes, M.; Zuo, X.N.; Gohel, S.; Kelly, C.; Smith, S.M.;
  Beckmann, C.F.; Adelstein, J.S.; Buckner, R.L.; Colcombe, S.;  et~al.
\newblock Toward discovery science of human brain function.
\newblock {\em PNAS} {\bf 2010}, {\em 107},~4734--4739.

\bibitem[Xu et~al.(2020)Xu, Liang, Ou, Li, Luo, and Tan]{xu2020sex}
Xu, M.; Liang, X.; Ou, J.; Li, H.; Luo, Y.j.; Tan, L.H.
\newblock Sex differences in functional brain networks for language.
\newblock {\em Cereb. Cortex} {\bf 2020}, {\em 30},~1528--1537.

\bibitem[Logan and Johnston(2010)]{logan2010investigating}
Logan, S.; Johnston, R.
\newblock Investigating gender differences in reading.
\newblock {\em Educ. Rev.} {\bf 2010}, {\em 62},~175--187.

\bibitem[Marrocco and McEwen(2016)]{marrocco2016sex}
Marrocco, J.; McEwen, B.S.
\newblock Sex in the brain: hormones and sex differences.
\newblock {\em Dialogues Clin. Neurosci.} {\bf 2016}, {\em 18},~373--383.

\bibitem[Lov{\'e}n et~al.(2011)Lov{\'e}n, Herlitz, and Rehnman]{loven2011women}
Lov{\'e}n, J.; Herlitz, A.; Rehnman, J.
\newblock Women’s own-gender bias in face recognition memory.
\newblock {\em Exp. Psychol.} {\bf 2011}.

\bibitem[Basso and Lowery(2004)]{basso2004global}
Basso, M.R.; Lowery, N.
\newblock Global-local visual biases correspond with visual-spatial
  orientation.
\newblock {\em J. Clin. Exp. Neuropsyc.} {\bf 2004}, {\em 26},~24--30.

\bibitem[Munion et~al.(2019)Munion, Stefanucci, Rovira, Squire, and
  Hendricks]{munion2019gender}
Munion, A.K.; Stefanucci, J.K.; Rovira, E.; Squire, P.; Hendricks, M.
\newblock Gender differences in spatial navigation: Characterizing wayfinding
  behaviors.
\newblock {\em Psychon. Bull. Rev.} {\bf 2019}, {\em 26},~1933--1940.

\bibitem[M{\"a}ntyl{\"a}(2013)]{mantyla2013gender}
M{\"a}ntyl{\"a}, T.
\newblock Gender differences in multitasking reflect spatial ability.
\newblock {\em Psychol. Sci.} {\bf 2013}, {\em 24},~514--520.

\bibitem[Generoso~Jr et~al.(2016)Generoso~Jr, Latoures, Acar, Miller, Ciano,
  Sandrei, Vieira, Luong, Hirsch, and Fidler]{generoso2016simulation}
Generoso~Jr, J.R.; Latoures, R.E.; Acar, Y.; Miller, D.S.; Ciano, M.; Sandrei,
  R.; Vieira, M.; Luong, S.; Hirsch, J.; Fidler, R.L.
\newblock Simulation training in early emergency response (steer).
\newblock {\em J. Contin. Educ. Nurs.} {\bf 2016}, {\em 47},~255--263.

\bibitem[Lurie et~al.(2020)Lurie, Kessler, Bassett, Betzel, Breakspear,
  Kheilholz, Kucyi, Li{\'e}geois, Lindquist, McIntosh,
  et~al.]{lurie2020questions}
Lurie, D.J.; Kessler, D.; Bassett, D.S.; Betzel, R.F.; Breakspear, M.;
  Kheilholz, S.; Kucyi, A.; Li{\'e}geois, R.; Lindquist, M.A.; McIntosh, A.R.;
  et~al.
\newblock Questions and controversies in the study of time-varying functional
  connectivity in resting fMRI.
\newblock {\em Netw. Neurosci.} {\bf 2020}, {\em 4},~30--69.

\bibitem[Cheng et~al.(2018)Cheng, Zhu, Sun, Deng, He, Yang, Ling, Ayaz, Fu, and
  Tong]{cheng2018principal}
Cheng, L.; Zhu, Y.; Sun, J.; Deng, L.; He, N.; Yang, Y.; Ling, H.; Ayaz, H.;
  Fu, Y.; Tong, S.
\newblock Principal states of dynamic functional connectivity reveal the link
  between resting-state and task-state brain: An fMRI study.
\newblock {\em Int. J. Neural Syst.} {\bf 2018}, {\em 28},~1850002.

\end{thebibliography}


%


\PublishersNote{}
\end{adjustwidth}
\end{document}